\documentclass[12pt]{article}

\begin{document}
\begin{center}
{\bf On the Generalized Dirac Equation for Fermions with Two Mass States}\\
\vspace{5mm}
 S. I. Kruglov \\
\vspace{5mm}
\textit{University of Toronto at Scarborough,\\ Physical and Environmental Sciences Department, \\
1265 Military Trail, Toronto, Ontario, Canada M1C 1A4}
\end{center}

\begin{abstract}
The generalized Dirac equation of the second order, describing
particles with spin 1/2 and two mass states, is analyzed. The
projection operators extracting states with definite energy and
spin projections are obtained. The first order generalized Dirac
equation in the 20-dimensional matrix form and the
relativistically invariant bilinear form are derived. We obtain
the canonical energy-momentum tensor and density of the
electromagnetic current expressed through the 20-component wave
function. Minimal and non-minimal electromagnetic interactions of
fermions are considered, and the quantum-mechanical Hamiltonian is
found. It is shown that there are only causal propagations of
waves in the approach considered.

\end{abstract}

\section{Introduction}

One of the modern problems is to explain the number of quark and
lepton generations, and their mass spectrum. The standard model
(SM) of electroweak interactions contains many free parameters:
$m_e$, $m_\mu$, $m_\tau$, $m_u$, $m_d$, $m_c$, $m_s$, $m_t$,
$m_b$, four mixing angles and three coupling constants. One of the
ways to reduce the number of free parameters is to explore the
Grand Unification Theories (GUTs) with different gauge groups. At
the GUTs scale, we have only one coupling constant, and masses of
leptons and quarks are generated below the GUTs scale by the
spontaneous gauge symmetry breaking with the help of the Higgs
mechanism. As a result of the symmetry breaking, in the framework
of the GUT, the generations of fermions appear. Nevertheless, the
Higgs Lagrangians also contain free parameters. Recent
observations of neutrino oscillations show that SM should be
extended, and possibly a fourth generation exists. The deeper
insight into the dynamics of the mass generation mechanism is
needed to understand physics beyond the SM.

We pay attention here on the Barut work \cite{Barut1}, (see also
\cite{Barut2}) who suggested a mass formula for $e$-,
$\mu$-leptons based on the generalized Dirac equation of the
second order (see \cite{Barut}, \cite{Wilson}) describing
particles with two mass states (see \cite{Kruglov3},
\cite{Kruglov4} for the case of bosonic fields with two mass
states). This equation may be considered as an effective one for
partly ``dressed" fermions, i.e. some radiation corrections are
taken into account. This scheme represents the non-perturbative
approach to quantum electrodynamics.

The goal of this paper is to formulate the mentioned second order
equation in the form of the first order generalized Dirac equation
(FOGDE), and investigate it.

The paper is organized as follows. In Sec. 2, we define the mass
spectrum of the generalized Dirac equation of the second order and
find the projection operator extracting the solutions in the
momentum space for free particles. The FOGDE is derived in Sec. 3.
The relativistically invariant bilinear form, the Lagrangian
formulation, and spin projection operators are given. In Sec. 4,
we obtain the canonical energy-momentum tensor and the
electromagnetic current density from the Lagrangian. Sec. 5 is
devoted to the introduction of minimal and non-minimal
electromagnetic interactions of fermions in the approach
discussed. The quantum-mechanical Hamiltonian is obtained. We make
a conclusion in Sec. 6. The system of units $\hbar =c=1$ is chosen
and notations as in \cite{Ahieser} are used.

\section{Spin-1/2 Field Equation of the Second Order}

Let us consider the second order (in derivatives) field equation
describing spin-1/2 particles introduced in \cite{Barut} (see also
generalizations in \cite{Dvoeglazov}):
\begin{equation}
\left(\alpha_1\gamma_\nu\partial_\nu+ \alpha_2\partial_\mu^2
+\kappa \right)\psi(x)=0 , \label{1}
\end{equation}
where $\partial_\nu =\partial/\partial x_\nu =(\partial/\partial
x_m,\partial/\partial (it))$, $\psi (x)$ is a bispinor. Repeated
indices imply a summation. The Dirac matrices $\gamma_\mu $ obey
the commutation relations $\gamma_\mu \gamma_\nu +\gamma_\nu
\gamma_\mu =2\delta_{\mu\nu}$. As physical values depend only on
two parameters, it is convenient to introduce new variables
$\kappa/\alpha_1=m$, $\alpha_2/\alpha_1=-a/m$, where $m$ is a
parameter with the dimension of the mass, and $a$ is a massless
parameter. With these arrangements, Eq. (1) is rewritten as
\begin{equation}
\left(\gamma_\nu\partial_\nu- \frac{a}{m}\partial_\mu^2 +
m\right)\psi(x)=0 . \label{2}
\end{equation}

In momentum space Eq. (2) for positive energy is given by
\begin{equation}
\left(i\widehat{p}+\frac{ap^2}{m}+m\right) \psi(p)=0  ,
 \label{3}
\end{equation}
where $\widehat{p}=\gamma_\mu p_\mu$. One has to make the
replacement $p\rightarrow -p$ when considering antiparticles.
Applying the operator $(- i \widehat{p} + a p^2 /m + m)$ to both
sides of Eq. (3),  we obtain the equation
\begin{equation}
\left[a^2p^4+m^2\left(2a+1\right)p^2+m^4\right]\psi(p)=0 ,
 \label{4}
\end{equation}
where $p^2= \textbf{p}^2-p_0^2$. In order to have a non-zero wave
function in Eq. (4), the diagonal elements need to be zero:
\begin{equation}
p^2=-m^2\left(\frac{2a+1 \pm \sqrt{4a+1}}{2a^2}\right) .
 \label{5}
\end{equation}
Eq. (5) defines the squared masses of fermions:
\begin{equation}
m_1^2=m^2\left(\frac{2a+1 - \sqrt{4a+1}}{2a^2}\right)
,~~m_2^2=m^2\left(\frac{2a+1 + \sqrt{4a+1}}{2a^2}\right) .
 \label{6}
\end{equation}
From Eqs. (6), one finds the restriction on the parameter $a$:
$a\geq -1/4$. At $a=-1/4$ both masses are equal, $m_1=m_2$. We
obtain from Eqs. (6) masses of spin-1/2 particles
\begin{equation}
m_1=-m\left(\frac{1 - \sqrt{4a+1}}{2a}\right)
,~~m_2=-m\left(\frac{1 +\sqrt{4a+1}}{2a}\right) .
 \label{7}
\end{equation}
In addition, we have a solution with opposite signs of masses. The
signs of masses in Eqs. (7) are chosen to be positive for the
negative parameter $a$. The negative mass solutions of Eqs. (6)
can be incorporated in a description of antiparticles. When the
parameter $a$ approaches to zero, and the parameter $m$ is fixed,
masses of the fermionic states, $m_1$, $m_2$, become unlimited.
Therefore, the limit $a=0$ should be made in Eqs (2), (3), and one
recovers the Dirac equation which describes fermions with the mass
$m$.

Eqs. (7) lead to the mass formula $m_1+m_2
=-m/a=\alpha_1/\alpha_2$ obtained in \cite{Barut}. The authors
\cite{Barut} interpreted the state of fermion with the mass $m_1$
as an electron, and the state of fermion with the mass $m_2$ as a
muon.

Now we construct the projection operators extracting solutions of
Eq. (3). It is easy to verify that projection operators (see the
general method of projection operators in \cite{Fedorov})
\begin{equation}
\Lambda_{\pm}= \frac{ap^2+m^2\mp
im\widehat{p}}{2\left(ap^2+m^2\right)} ,\label{8}
\end{equation}
which obey the necessary equation $\Lambda_{\pm}^2=
\Lambda_{\pm}$, are solutions of equations
\[
\left(\pm i\widehat{p}+\frac{ap^2}{m}+m\right) \Lambda_{\pm}=0 .
\]
The projection operator $\Lambda_{+}$ corresponds to positive
energy, and $\Lambda_{-}$ - to negative energy of particles.

\section{First Order Generalized Dirac Equation}

In order to formulate the FOGDE, we introduce the 20-dimensional
function
\begin{equation}
\Psi (x)=\left\{ \psi _A(x)\right\} =\left(
\begin{array}{c}
\psi (x)\\
\psi _\mu (x)
\end{array}
\right) \hspace{0.5in}(\psi_\mu (x)=-\frac{1}{m}\partial_\mu \psi
(x)) , \label{9}
\end{equation}
where $A=0,\mu$. The function $\Psi (x)$ in Eq. (9) is the direct
sum of a bispinor $\psi (x)$ and a vector-bispinor $\psi _\mu
(x)$. Therefore, the wave function $\Psi (x)$ realizes the
reducible representation
$[(1/2,0)\oplus(0,1/2)]\oplus\{(1/2,1/2)\otimes
[(1/2,0)\oplus(0,1/2)]\}$ of the Lorentz group.

Let us introduce the elements of the entire algebra $\varepsilon
^{A,B}$ \cite{Bogush} with the properties
\begin{equation}
\left( \varepsilon ^{M,N}\right) _{AB}=\delta _{MA}\delta _{N,B},
\hspace{0.5in}\varepsilon ^{M,A}\varepsilon ^{B,N}=\delta
_{AB}\varepsilon ^{M,N}, \label{10}
\end{equation}
$A,B,M,N=0,1,2,3,4$. The $\varepsilon ^{M,N}$ represent 25 basis
elements corresponding to the different values of the superscripts
$M$, $N$, and they are 5x5 matrices whose elements are labelled by
the subscripts. The elements of the matrix $\varepsilon ^{M,N}$
consist of zeros and only one element is unity, where row $M$ and
column $N$ cross.

With the help of Eqs. (9), (10), Eq. (2) can be rewritten in the
form of the first order equation
\begin{equation}
\partial _\nu \left(\varepsilon ^{\nu,0 }+ a\varepsilon ^{0,\nu} +
\varepsilon^{0,0}\gamma_\nu\right)_{AB}\Psi _B(x)+ m\left[
\varepsilon ^{\mu ,\mu }+ \varepsilon ^{0,0}\right] _{AB}\Psi
_B(x)=0 , \label{11}
\end{equation}
where we imply that $\gamma$-matrices act on the bispinor
subspace. All repeated indices such as in Eq. (11) imply a
summation even in the superscripts of the elements of the algebra.
After introducing 20-dimensional matrices
\begin{equation}
\alpha _\nu =\left(\varepsilon ^{\nu,0 }+ a\varepsilon ^{0,\nu
}\right)\otimes I_4 + \varepsilon^{0,0}\otimes\gamma_\nu ,
\label{12}
\end{equation}
where unit four-dimensional matrix $I_4$ acts on bispinor subspace
and unit five-dimensional matrix $1\equiv I_5=\varepsilon ^{\mu
,\mu }+ \varepsilon ^{0,0}$ acts on scalar-vector subspace, Eq.
(11) takes the form of the FOGDE:
\begin{equation}
\left( \alpha _\nu \partial _\nu +m\right) \Psi (x)=0 . \label{13}
\end{equation}
Eq. (13) is convenient for investigations of spin-1/2 fermions
with two mass states. According to the general theory
\cite{Gel'fand}, the masses of particles described by the first
order relativistic wave equation (13) are $m/\lambda_i$, where
$\lambda_i$ are the eigenvalues of the matrix $\alpha_4$. In our
case the matrix $\alpha _4 =\left(\varepsilon ^{4,0 }+
a\varepsilon ^{0,4 }\right)\otimes I_4 +
\varepsilon^{0,0}\otimes\gamma_4$ obeys the minimal matrix
equation
\begin{equation}
 \alpha _4^4-(1+2a)\alpha_4^2 +a^2 =0 . \label{14}
\end{equation}
One can obtain from Eq. (14) four eigenvalues of the matrix
$\alpha_4$:
\begin{equation}
\pm\lambda_1,~~\pm\lambda_2,~~~~\lambda_1=\frac{-1 -
\sqrt{4a+1}}{2},~~\lambda_2=\frac{-1 + \sqrt{4a+1}}{2} .
\label{15}
\end{equation}
It is easy to verify that masses of fermion states of
20-dimensional matrix Eq. (13): $m_1=m/\lambda_1$,
$m_2=m/\lambda_2$ coincide with those in Eq. (7). Two additional
masses possess opposite signs.

To build the spin operators, one needs the generators of the
Lorentz group in the 20-dimensional representation. Such
generators are given by \cite{Kruglov}, \cite{Kruglov1}
\begin{equation}
J_{\mu \nu }=J_{\mu \nu }^{(1)}\otimes I_4+I_5 \otimes
J_{\mu\nu}^{(1/2)} ,
 \label{16}
\end{equation}
\begin{equation}
J_{\mu\nu}^{(1)}= \varepsilon^{\mu,\nu}-\varepsilon^{\nu,\mu} ,
 \label{17}
\end{equation}
\begin{equation}
J_{\mu\nu}^{(1/2)}= \frac{1}{4}\left( \gamma_\mu
\gamma_\nu-\gamma_\nu \gamma_\mu \right) .
 \label{18}
\end{equation}
Operators $J_{\mu\nu}^{(1)}$, $J_{\mu\nu}^{(1/2)}$ are the
generators of the Lorentz group in four-dimensional vector and
bispinor spaces, respectively. Generators (16)-(18) obey the
commutation relations
\begin{equation}
\left[ J_{\mu \nu },J_{\alpha \beta}\right] =\delta _{\nu \alpha
}J_{\mu \beta}+\delta _{\mu \beta }J_{\nu \alpha}-\delta _{\nu
\beta }J_{\mu \alpha}-\delta _{\mu \alpha }J_{\nu \beta} .
\label{19}
\end{equation}
The antisymmetric parameters of the Lorentz group $\omega_{mn}$
($m,n=1,2,3$) are real, and $\omega_{m4}$ are imaginary in our
metric. The commutation relation
\begin{equation}
\left[ \alpha _\lambda ,J_{\mu \nu }\right] =\delta _{\lambda \mu
}\alpha _\nu -\delta _{\lambda \nu }\alpha _\mu   \label{20}
\end{equation}
holds and guarantees the relativistic form-invariance of Eq. (13).

Now we consider operators of the spin projections on the direction
of the momentum $\textbf{p}$:
\begin{equation}
\sigma_p=-\frac{i}{2|\textbf{p}|}\epsilon_{abc}\textbf{p}_a J_{bc}
, \label{21}
\end{equation}
where $|\textbf{p}| =\sqrt{p_1^2 +p_2^2+p_3^2}$. Using Eq. (20),
it is not difficult to prove that the operator of Eq. (13) in the
momentum space $(i\alpha_\mu p_\mu +m)$ commutes with the spin
operator (21): $[i\alpha_\mu p_\mu +m,\sigma_p]=0$. Therefore, the
operators $(i\alpha_\mu p_\mu +m)$ and $\sigma_p$ have the common
eigenfunction in the momentum space. It is not difficult to verify
that the minimal matrix equation
\[
\left(\sigma_p^2-\frac{1}{4}\right)\left(\sigma_p^2-\frac{9}{4}\right)=0
\]
holds. With the aid of the method \cite{Fedorov}, we find the
projection operators
\begin{equation}
P_{\pm 1/2}=\mp\frac{1}{2}\left(\sigma_p\pm\frac{1}{2}
\right)\left(\sigma_p^2-\frac{9}{4}\right)  \label{22}
\end{equation}
which extract spin projections $\pm 1/2$, so that
\[
\sigma_p P_{\pm 1/2}=\pm \frac{1}{2} P_{\pm 1/2} .
\]
The operators $P_{\pm 1/2}$ acting on the arbitrary
$20$-dimensional function $\Psi_0 (p)$ produce the eigenfunction
$\Psi_{\pm 1/2} (p)=P_{\pm 1/2}\Psi_0 (p)$ of the operator
$\sigma_p$: $\sigma_p \Psi_{\pm 1/2}=\pm (1/2)\Psi_{\pm 1/2}$.

To obtain the Lagrangian, one has to find the Hermitianizing
matrix $\eta$ in our 20-dimensional representation space which
obeys the relations \cite{Gel'fand}
\begin{equation}
\eta \alpha _m=-\alpha _m^{+}\eta ,\hspace{0.5in}\eta \alpha
_4=\alpha _4^{+}\eta \hspace{0.5in}(m=1,2,3) .  \label{23}
\end{equation}
Such a matrix exists, and is given by
\begin{equation}
\eta=\left(a\varepsilon^{m,m}-a\varepsilon^{4,4}-
\varepsilon^{0,0}\right)\otimes\gamma_4 . \label{24}
\end{equation}

Using Eqs. (23), we obtain from Eq. (13) the equation
\begin{equation}
\overline{\Psi }(x)\left( \alpha _\mu \overleftarrow{\partial}
_\mu -m\right) =0 , \label{25}
\end{equation}
where $\overline{\Psi }(x)=\Psi ^{+}(x)\eta$, and $\Psi ^{+}(x)$
is the Hermitian-conjugate wave function. The derivative
$\overleftarrow{\partial} _\mu$ in Eq. (25) acts on the
left-standing function. The relativistically invariant bilinear
form is given by $ \overline{\Psi }(x)\Psi (x)=\Psi ^{+}(x)\eta
\Psi (x)$. Thus, the Lagrangian reads:
\begin{equation}
{\cal L}=-\frac{1}{2}\left[\overline{\Psi }(x)\left(\alpha _\mu
\partial _\mu +m\right) \Psi (x)-\overline{\Psi }(x)\left(\alpha _\mu
\overleftarrow{\partial} _\mu -m\right) \Psi (x)\right] .
\label{26}
\end{equation}
Variation of the action $S=\int {\cal L}d^4x$ with respect to the
independent fields $\Psi (x)$, $\overline{\Psi }(x)$,
corresponding to Lagrangian (26), gives Euler-Lagrange equations
(13), (25).

\section{The Energy-Momentum Tensor and Electromagnetic Current}

The energy-momentum tensor can be found with the help of the
standard procedure, using the general formula \cite{Bogolubov}
\begin{equation}
T_{\mu\nu}=\frac{\partial\mathcal{L}}{\partial\left(\partial_\mu
\Psi (x)\right)}\partial_\nu \Psi (x)+\partial_\nu \overline{\Psi
}(x) \frac{\partial\mathcal{L}}{\partial\left(\partial_\mu
\overline{\Psi} (x)\right)}-\delta_{\mu\nu} \mathcal{L} .
\label{27}
\end{equation}
With the help of Eq. (27), one may obtain from the Lagrangian (26)
the canonical energy-momentum tensor as follows
\begin{equation}
T_{\mu\nu}=\frac{1}{2}\left(\partial_\nu \overline{\Psi}
(x)\right)\alpha_\mu \Psi (x)-\frac{1}{2} \overline{\Psi}
(x)\alpha_\mu \partial_\nu\Psi (x) . \label{28}
\end{equation}
We took here into consideration that $ \mathcal{L}=0$ for
functions $\Psi (x)$, $\overline{\Psi} (x)$ obeying Eqs. (13),
(25). Canonical energy-momentum tensor (28) is conserved but is
not symmetric. It can be symmetrized \cite{Wilson} by adding the
divergence term to the Lagrangian (26) which does not change
Euler-Lagrange equations (13), (25). With the help of Eqs. (9),
(12), one may obtain from Eq. (28) the expression
\[
T_{\mu\nu}=\frac{1}{2}\overline{\psi} (x)\gamma_\mu
\partial_\nu\psi (x)-\frac{1}{2}\left(\partial_\nu\overline{\psi} (x)\right)\gamma_\mu
\psi (x)+\frac{a}{2m}\left(\partial_\mu\overline{\psi}
(x)\right)\partial_\nu \psi (x)
\]
\[
-\frac{a}{2m}\left(\partial_\mu\partial_\nu \overline{\psi}
(x)\right)\psi (x)+ \frac{a}{2m}\left(\partial_\nu \overline{\psi}
(x)\right)\partial_\mu\psi (x) -\frac{a}{2m}\overline{\psi}
(x)\partial_\nu
\partial_\mu\psi (x) .
\]
The first two terms in this equation correspond to the ordinary
Dirac equation. The energy density and the momentum density are
given by ${\cal E}=T_{44}$, $P_m=iT_{m4}$.

The conserved electric current density may be obtained by the
relationship \cite{Bogolubov}
\begin{equation}
j_\mu (x)=i\left( \overline{\Psi }(x)
\frac{\partial\mathcal{L}}{\partial\left(\partial_\mu
\overline{\Psi} (x)\right)}-
\frac{\partial\mathcal{L}}{\partial\left(\partial_\mu \Psi
(x)\right)}\Psi (x)\right) . \label{29}
\end{equation}
Substituting Eq. (26) into Eq. (29), we get the electric current
density
\begin{equation}
j_\mu (x)=i\overline{\Psi }(x)\alpha_\mu \Psi(x) . \label{30}
\end{equation}
From Eqs. (13), (25), one may prove the conservation of the
four-vector current density: $\partial_\mu j_\mu (x)=0$. The
canonical energy-momentum tensor (28) and electric current density
(30) are expressed through 20-component wave function $\Psi (x)$.
Using Eqs. (9), (12), we  obtain from Eq. (30) the electric
current density only in terms of the bispinor $\psi (x)$:
\[
j_\mu (x)=-i\overline{\psi }(x)\gamma_\mu \psi(x)+
\frac{ia}{m}\overline{\psi }(x)\partial_\mu \psi(x)
-\frac{ia}{m}\left(\partial_\mu\overline{\psi}(x)\right) \psi(x).
\]
It follows from this expression that the current (30) is related
to Barut's original anomalous magnetism. So, the $j_\mu (x)$
includes the usual Dirac current as well as convective terms.

\section{Electromagnetic Interactions of Fermions}

It follows from Eq. (14) that the inverse matrix $\alpha_4^{-1}$
exists, as there are no zero eigenvalues of the matrix $\alpha_4$.
One may find from Eq. (14) the inverse matrix
\begin{equation}
\alpha_4^{-1}=\frac{2a+1}{a^2}\alpha_4 -\frac{\alpha_4^3}{a^2} .
\label{31}
\end{equation}
This indicates that all components of the wave function $\Psi
(x)$, Eq. (9), are canonical, and contain time derivatives. Thus,
there are no subsidiary conditions in the first order wave
equation (13), and it can be rewritten in the Hamiltonian form.
The minimal interaction with electromagnetic field in the first
order equation (13) can be obtained by the substitution $\partial
_\mu \rightarrow D_\mu =\partial _\mu -ieA_\mu $ ($A_\mu $ is the
four-vector potential of the electromagnetic field). We also
introduce non-minimal interaction by adding terms linear in the
strength of the electromagnetic field $\mathcal{F}_{\mu \nu
}=\partial _\mu A_\nu -\partial _\nu A_\mu $. So, we postulate the
matrix equation
\begin{equation}
\biggl [\alpha _\mu D_\mu +\frac i2\left( \kappa _0P_0+\kappa _1
P_1\right) \alpha _{\mu \nu }\mathcal{F}_{\mu \nu }+m\biggr ]\Psi
(x)=0  ,\label{32}
\end{equation}
where the projection operators $P_0$, $P_1$, are given by
\begin{equation}
P_0=\varepsilon ^{0,0}\otimes I_4,\hspace{0.3in}P_1=\varepsilon
^{\mu ,\mu }\otimes I_4 ,
 \label{33}
\end{equation}
and
\[
\alpha _{\mu \nu }=\alpha _\mu \alpha _\nu -\alpha _\nu \alpha
_\mu
\]
\begin{equation}
=\varepsilon^{0,0}\otimes\left(\gamma_\mu\gamma_\nu-\gamma_\nu\gamma_\mu\right)
+\varepsilon^{\mu,0}\otimes\gamma_\nu-\varepsilon^{\nu,0}\otimes\gamma_\mu\
\label{34}
\end{equation}
\[
+\varepsilon^{0,\nu}\otimes\gamma_\mu-\varepsilon^{0,\mu}\otimes\gamma_\nu\
+a\left(\varepsilon^{\mu,\nu}-\varepsilon^{\nu,\mu}\right)\otimes
I_4.
\]
The projection operators $P_0$, $P_1$ extract scalar, vector
subspaces and obey the relations: $P_0^2=P_0$, $P_1^2=P_1$,
$P_0+P_1=1$. It is easy to verify that Eq. (32) is form-invariant
under the Lorentz transformations. We have introduced two
parameters $\kappa_0$, $\kappa_1$ which characterize anomalous
electromagnetic interactions of fermions. It should be noted that
the relativistic invariance allows us to introduce only two
parameters because there are two subspaces corresponding to the
projection operators $P_0$ and $P_1$.

With the help of Eqs. (9), (10, (12), we obtain the tensor form of
Eq. (32):
\begin{equation}
\left( \gamma_\nu D_\nu +i\kappa_0 \gamma_\mu \gamma_\nu
\mathcal{F}_{\mu \nu} + m\right)\psi (x)+\left(aD_\mu
+i\kappa_0\gamma_\nu\mathcal{F}_{\nu \mu}\right)\psi_\mu (x)=0 ,
 \label{35}
\end{equation}
\begin{equation}
\left(D _\mu +i\kappa_1\gamma_\nu\mathcal{F}_{\mu \nu}\right) \psi
(x)+ \left(m\delta_{\mu\nu}+i\kappa_1 a\mathcal{F}_{\mu
\nu}\right)\psi_\nu (x)=0 .
 \label{36}
\end{equation}
It should be stressed that Eq. (36) is an analog of the gradient
equation (see Eq. (9)) in the case of interacting fields. System
of Eqs. (35), (36) defines  bispinor $\psi (x)$ as well as
vector-bispinor $\psi_\nu (x)$. Eq. (36) may be considered as a
matrix equation for vector-bispinor $\psi_\nu (x)$. After finding
the inverse matrix $\left(m\delta_{\mu\nu}+i\kappa_1
a\mathcal{F}_{\mu \nu}\right)^{-1}$, one can express the $\psi_\nu
(x)$ from Eq. (36) and replace into Eq. (35) to get an equation
for bispinor $\psi (x)$ describing the interaction of fermions
with electromagnetic fields. This equation includes the
interaction with the anomalous magnetic moment of particles. Eqs.
(35), (36) can also be applied for phenomenological descriptions
of composite fermions.

Let us find the quantum-mechanical Hamiltonian from Eq. (32). Eq.
(32) can be rearranged as
\[
i\alpha _4\partial _t\Psi (x)=\biggl
[\alpha _aD_a+m+eA_0\alpha _4+
\]
\begin{equation}
+\frac{i}{2}\left( \kappa _0 P_0+\kappa _1P_1 \right) \alpha _{\mu
\nu }\mathcal{F}_{\mu \nu }\biggr ]\Psi (x).  \label{37}
\end{equation}
One can obtain from Eq. (37) the Hamiltonian form of the equation
\[
i\partial _t\Psi (x)=\mathcal{H}\Psi (x) ,
\]
\vspace{-7mm}
\begin{equation} \label{38}
\end{equation}
\vspace{-7mm}
\[
\mathcal{H}=\alpha _4^{-1}\biggl [\alpha _aD_a+m+eA_0\alpha _4+
\frac i2\left( \kappa _0P_0+\kappa _1P_1 \right) \alpha _{\mu \nu
}\mathcal{F}_{\mu \nu }\biggr ] ,
\]
where the matrix $\alpha_4^{-1}$ is given by Eq. (31).

To investigate the consistency problem, in accordance with the
method \cite{Zwanziger}, we replace the derivatives in Eq. (32) by
the four-vector $n_\mu$ to get the characteristic surfaces. One
can obtain the normals to the characteristic surfaces, $n_\mu$, by
solving the equation
\begin{equation}
 \det\left(\alpha_\mu n_\mu\right) =0 . \label{39}
\end{equation}
Using the frame of reference where $n_\mu=(0,0,0,n_4)$, Eq. (39)
becomes $\det(\alpha_4 n_4)=0$. It follows from Eq. (14) that
there are no zero eigenvalues of the matrix $\alpha_4$. Therefore,
equation $\det(\alpha_4 n_4)=0$ possesses only the trivial
solution $n_4=0$, that indicates: there are only causal
propagations of waves as in the Dirac equation, and no
superluminal speed of waves.

\section{Conclusion}

The projection operators extracting states with definite energy
and spin projections were constructed for the generalized Dirac
equation of the second order, describing particles with spin 1/2
and two mass states. We have obtained the FOGDE in the
20-dimensional matrix form, the relativistically invariant
bilinear form, and the Lagrangian. The FOGDE is convenient for
different applications. The canonical energy-momentum tensor and
density of the electromagnetic current have expressed through the
20-component wave function. We have introduced two parameters
characterizing non-minimal electromagnetic interactions of
fermions. As a particular case, they include the interaction of
the anomalous magnetic moment of particles. The quantum-mechanical
Hamiltonian has obtained and the Hamiltonian form of the equation
was given. It was shown that there are only causal propagations of
waves in the approach considered.

In order to get the third generation of fermions, one needs to
introduce a third-order (in derivatives) generalized Dirac
equation. Using the technique described, such equation can be
reduced to first-order equation such as Eq. (13). There is not a
natural limit to these procedures, corresponding to a finite
number of generations in the approach considered.

The scheme investigated can be applied for a consideration of the
two flavour generations of leptons and quarks.  Possibly, Eqs.
(35), (36) may be considered as effective equations for
electromagnetic interactions of hadrons.

\end{document}